\documentclass[a4paper, 12pt]{article}

\usepackage{epsfig,amssymb,euscript,xspace, color}
\usepackage{amsmath,jheppub_0,mathtools}
\def\bea{\begin{eqnarray}}
\def\eea{\end{eqnarray}}
\def\be{\begin{equation}}
\def\ee{\end{equation}}

\newcommand{\rpl}{r_{\!_+}}
\newcommand{\rmi}{r_{\!_-}}



\begin{document}

\title{Fast Scrambling due to Rotating Shockwaves in BTZ  }
\author{Vinay Malvimat$^a$, Rohan R. Poojary$^b$}
\affiliation{$^a$Theory Division, Saha Institute of Nuclear Physics, \\
Homi Bhaba National Institute (HBNI), \\
1/AF, Bidhannagar, Kolkata 700064, India. 
\vspace{0.3cm}
\\
$^b$Institute for Theoretical Physics, TU Wien,\\
Wiedner Hauptstrasse 8-10, 1040 Vienna, Austria.  }
\emailAdd{vinaymmp@gmail.com, rpglaznos@gmail.com}

\abstract{We study the perturbation due to rotating shockwaves in BTZ geometries at late times and analyse the change in Mutual Information between the two subsystems belonging to the dual CFT$_L$ and CFT$_R$. We find that the scrambling of Mutual Information is in general governed by the Lyapunov index $\lambda_L$ which is bounded by $\kappa=\frac{2\pi}{\beta(1-\mu\mathcal{L})}\geq \frac{2\pi}{\beta}$ where $\mu=\rmi/\rpl$ and $\mathcal{L}$ is the angular momentum of the shockwave. For the special case of $\mathcal{L}$=$1$ we find the Mutual Information analytically and show that it is characterized by $\lambda_L=\kappa/2$  with  the scrambling time for large black holes given as $t_*=\frac{\beta(1-\mu)}{\pi}\log S$. }

\maketitle



\section{Introduction}
In recent times quantum chaos has been used as an important tool to understand novel qualitative features in quantum gravity $via$ holography. The phenomena of quantum chaos emerges in the study of how initial perturbations grow and disrupt entanglement between different subsystems of a large $N$ thermal quantum system \cite{Hayden:2007cs}. This is reflected in what is known as the ``scrambling time" $t_*$ in which the Mutual Information between two sub-systems is disrupted substantially due to a perturbation of few degrees of freedom. In systems which exhibit fast scrambling, the rate of growth of this perturbation is exponential, with black holes being amongst the fastest scramblers of this early perturbation \cite{Sekino:2008he,Susskind:2011ap,Lashkari:2011yi,Kitaev:2014talk}. The index of this exponential behaviour in time is termed as the Lyapunov index $\lambda_L$ and in large $N$ thermal systems $\lambda_L$ is found to be bounded by the temperature of the system \cite{Maldacena:2015waa}. This bound constrains  the maximum value that $\lambda_L$  can attain at any instant of time as the perturbation grows in the system, while the eventual growth of the perturbation until the scrambling time depends on observable which is being studied. One usually computes $\lambda_L$ by computing the Out of Time Ordered Correlators (OTOCs) \cite{Maldacena:2015waa,Halder:2019ric} or by computing the perturbations in the mutual information $I(A:B)$ between two subsystems in the theory \cite{Shenker:2013pqa,Caputa:2015waa,Stikonas:2018ane,Shenker:2013yza}. It is important to note that the later provides an upper bound on the correlators of the theory \cite{Wolf:2007tdq}\footnote{This statement is only clear in finite number of dimensions. We would  only be concerning ourselves with CFT$_2$ in this paper.}. It was found that for the case of Schwarzchild black holes in $AdS_3$ that $\lambda_L$ is indeed the temperature of the black hole \cite{Shenker:2013pqa,Shenker:2014cwa} thus validating the conjecture that black holes are among the fastest scramblers of in fallen information. This along with the study of simple 1d strongly coupled solvable SYK model \cite{Maldacena:2016hyu,Polchinski:2016xgd} and JT gravity as the near horizon effective gravitational action for near extremal black holes \cite{Maldacena:2016upp,Jensen:2016pah} has led to a flurry of activity especially in understanding the entropic black hole information paradox and contributions from wormholes to the gravity path integral $c.f.$ \cite{Penington:2019kki,Almheiri:2019qdq,Penington:2019npb,Marolf:2020xie}.
\\\\
Rotating black hole geometries in $AdS$ correspond to a boundary CFT with a fixed temperature $T_H$ and chemical potential $\mu$, thus the bound investigated in \cite{Maldacena:2015waa} does not apply.
For the case of rotating black holes in $AdS_3$ it was found that $\lambda_L$ can be the greater of the two temperatures in the CFT$_2$ \cite{Stikonas:2018ane,Poojary:2018eszz,Jahnke:2019gxr}. This also lead to an analysis \cite{Halder:2019ric} of the arguments in \cite{Maldacena:2015waa} in the presence of chemical potential $\mu$ for globally conserved charge which also found that the instantaneous Lyapunov exponent to be bounded by 
\be
\lambda_L\leq \frac{2\pi}{\beta(1-\mu/\mu_c)},
\label{Halder_bound}
\ee
where $\mu_c$ is the maximum value attainable by $\mu$. 
It was additionally found \cite{Mezei:2019dfv,Craps:2020ahu,Craps:2021bmz} that the OTOC in rotating BTZ had a scrambling time governed by smaller of the two temperatures of the CFT$_2$ with the growth in OTOC exhibiting a sawtooth pattern where for small time scales $\lambda_L$ was found to be greater of the two temperatures. This doesn't seem to contradict the results in  \cite{Halder:2019ric} as the arguments of \cite{Maldacena:2015waa,Halder:2019ric} only constrain $\lambda_L$ and not it's average, which may govern the scrambling time. Scrambling time for large (fast scrambling) systems at temperature $T_H$ typically is expected to scale with the $d.o.f.$ of the system $i.e.$ $t_*\sim \tfrac{1}{T_H}\log S$ which can be parametrically larger than the  dissipation time of the system $t_{\rm dis}=T_H^{-1}$. Therefore simply determining the instantaneous value of $\lambda_L(\sim T_H)$ can only constrain $t_*$ from above.    
\\\\
The computation of OTOCs in the bulk essentially involves knowing the bulk to boundary propagators in a desired black hole background and therefore is only feasible in locally $AdS_3$ geometries \cite{Shenker:2014cwa, Blake:2021hjj}. On the other hand measuring scrambling via disruption of Mutual Information between two subsystems in the left and right CFTs in a TFD state involves computing relevant Ryu-Takayanagi \cite{Ryu:2006bv,Ryu:2006ef} or the covariantized HRT surfaces \cite{Hubeny:2007xt} \cite{Shenker:2013pqa,Caputa:2015waa,Stikonas:2018ane}. See also \cite{Maldacena:2001kr,Hartman:2013qma,Caputa:2013eka,Hubeny:2007xt,Mandal:2014wfa} for the use of RT and HRT surfaces in holography. For the case of rotating BTZ this was essentially done in the bulk in \cite{Stikonas:2018ane} by utilising embedding coordinates to write down a perturbed BTZ background.
%
%
This method cannot be utilised in higher dimensions as once again only global  $AdS$ (or patches of it) can be described in terms of embedding coordinates. It would be useful to understand the results of \cite{Stikonas:2018ane,Poojary:2018eszz,Jahnke:2019gxr} and \cite{Mezei:2019dfv,Craps:2020ahu,Craps:2021bmz} in terms of the setup described in \cite{Shenker:2013pqa} which uses a shockwave to perturb a Schwarzchild black hole. This setup has the advantage of understanding the Lyapunov index  due to the blueshift suffered by the in-falling shockwave and the subsequent change in the black hole's geometry at late times $via$ knowing the Dray-'t Hooft solution \cite{Sfetsos:1994xa,Dray:1984ha}. 
\\\\
We investigate how a generic BTZ background can be perturbed by a shockwave originating in the past from one of the boundaries with a non-zero angular momentum $\mathcal{L}$. We then compute the disruption of Mutual Information $I(A:B)$ between two intervals to discern the Lyapunov index and scrambling time when analytical solutions are possible. The essential physics of scrambling can be explained by the blueshift of the in falling quanta of energy $E\sim E_0 e^{\kappa t_0}$ as seen by the boundary observer at late times at $t\gg t_0$. As this perturbation grows it disrupts the fine tuned entanglement of the left and right CFTs specifying the TFD state. For static geometries $\kappa=\frac{2\pi}{\beta}$ and does not depend on the angular momentum of the perturbation\footnote{Here we are assuming that the angular momentum is  small enough for the perturbation to fall into the black hole.}. However for rotating geometries we find 
\be
\kappa=\frac{2\pi}{\beta(1-\mu\mathcal{L})}
\ee
which essentially depends on both the angular momentum $\mathcal{L}$ (per unit energy) of the perturbation and the horizon velocity $\mu=\rmi/\rpl$  of the black hole. This similar to the bound proposed in \cite{Halder:2019ric} $i.e.$ the $rhs$ of \eqref{Halder_bound}. It is known that $\mathcal{L}$  has to lie within a range determined by the black hole's $M,J$ in order to start at the boundary and end up in its interior. For the case of BTZ this range is given by \cite{Cruz:1994ir}
\be\frac{2\rpl\rmi}{\rpl^2+\rmi^2}\leq\mathcal{L}\leq 1.
\ee    
The dependence of $\kappa$ or the blueshift suffered by an in falling quanta of energy with angular momenta $\mathcal{L}$ in rotating geometries has not been investigated in the context of chaos in black holes.
We find that for the case of $\mathcal{L}=1$ the scrambling time is governed by a $\lambda_L=\kappa/2$ where $\kappa=\frac{2\pi}{\beta(1-\mu)}$ with $\mu=\rmi/\rpl$. Thus implying that for $\mu>1/2$ the disruption of mutual information happens at a rate governed by $\lambda_L>\frac{2\pi}{\beta}$ even for late times.
The methods of \cite{Shenker:2013pqa} were applied for the case of rotating BTZ in \cite{Reynolds:2016pmi} where it was concluded that the scrambling time for rotating black holes is governed by an exponent  $\lambda_L=T_H$ as in \cite{Shenker:2013pqa}. This result can be regarded as a special case of our result for $\mathcal{L}=0$. However we do comment on the  contrasting technique used here for computing the shockwave solutions in rotating geometries in subsection 2.2.  The equivalent CFT$_2$ computation in \cite{Stikonas:2018ane} determines both scrambling time and $\lambda_L$ to be controlled by smaller of the 2 temperatures in CFT$_2$, however one can easily observe that it could also have been controlled by larger of the 2 temperatures if the location of perturbation were change $w.r.t.$ the entangling surface $c.f.$ subsection 2.2 of this paper.     
\\\\
The analysis of the OTOCs in rotating BTZ \cite{Mezei:2019dfv,Craps:2020ahu} and its CFT dual \cite{Craps:2021bmz} essentially find that scrambling time is governed by the smaller of the two temperatures associated with the boundary CFT even though the instantaneous $\lambda_L$ could be the greater of the two temperatures. In contrast, we find that if one were to measure the scrambling of Mutual Information between large enough subsystems belonging to the left and the right CFTs of the TFD state by rotating shockwaves then the scrambling time is clearly found to be greater than the temperature of the black hole. This does not seem to contradict any of the computations in literature measuring OTOCs as measurement of entanglement entropy is non-local in its nature. We find that our results are consistent with the bound \eqref{Halder_bound} found in \cite{Halder:2019ric}.
\\\\
The paper is summarized as follows: In section 2 we review the essential setup of the computation by Shenker and Stanford \cite{Shenker:2013pqa} and also briefly review the result of \cite{Reynolds:2016pmi} which attempts to do the same in rotating BTZ. In section 3 we construct Kruskal coordinates along null geodesics with non-zero angular momentum $\mathcal{L}$ and find that the blueshift at the outer horizon at $t=0$ associated to the in-falling null particle released in the far past at $t_0$ from the boundary is 
\be
E\sim E_0e^{\kappa t_0},\,\,\,\,\kappa=\frac{2\pi}{\beta(1-\mu\mathcal{L})}\geq\frac{2\pi}{\beta}
\ee            
where $\mu=\rmi/\rpl$ and $1\geq\mathcal{L}\geq2\mu/(1+\mu^2)$. We also construct the Dray-t'Hooft solution due to such a shockwave resulting from  an axisymmetric shell of null particles We analyse the case for a single null particle in an appendix.
In Section 4 we compute and analyse the change in $I(A:B)$ due to the shell of null particles at late times and find that in general $\lambda_L<\kappa$. We are able to analytically solve for $I(A:B)$ for the case $\mathcal{L}=1$ where in we find $\lambda_L=\kappa/2$, which can be greater than the black holes temperature when $\mu>1/2$. As analytical solutions are obtained we also see that the scrambling time is indeed  given by $t_*\sim\frac{2}{\kappa}\log S$. 
This also includes extremal case of $\mu=1$. We plot $I(A:B)$ numerically to show that for certain cases with $2\pi/\beta<\kappa/2<\lambda_L<\kappa$. We conclude with discussions in section 5. 
\section{Non-Rotating Shockwaves in $AdS_3$}
The shock-wave computation of \cite{Shenker:2013pqa} can be easily summarised for the case of non rotating BTZ. The basic idea is to imagine the Kruskal extension of the eternal black hole with the 2 boundaries forming the thermofield double(TFD) (CFT$_L$ $\otimes$ CFT$_R$) state. The black hole in the bulk is an entangled TFD state of the boundary CFT. This entanglement can be measured by the mutual information between any two subregions of the boundary CFT (each for CFT$_L$ and CFT$_R$) by using the Ryu-Takayangi prescription. This basically implies measuring the extremal area  of a co-dimension 2 surface homologous to the two boundary subregions. When the sub-regions are large enough,  the extremal surface in the bulk traverses from one boundary to the other. The Shenker-Stanford computation then simply measures the time taken by a shockwave  emanating in the past to grow in time in the bulk and perturb this minimal area. 
\\\\
The computation then can roughly be demarcated into two parts: Computation of the metric in response to the shockwave set up in the distant past, and computing the geodesic distances in non-rotating BTZ with and without the shockwave back ground. The Lyapunov index $\lambda_L$ is then read off by observing the growth of this perturbation as it is regarded to scramble the  entanglement between the two CFTs.  We begin by expressing the Schwarzchild metric in Kruskal coordinates which are suitable for describing in-falling (out going) null trajectories
\bea
\frac{ds^2}{l^2}&=&\frac{dr^2}{f(r)}-f(r)dt^2+r^2d\phi^2,\hspace{0.3cm} w/\,\, f(r)=r^2-r_+^2\cr
&=&\frac{-4dUdV+r_+^2(1-UV)^2d\phi^2}{(1+U V)^2}\hspace{0.4cm}\cr&&\cr
w/&&U=-e^{-\kappa u}, V=e^{\kappa v}\hspace{0.4cm}\&\hspace{0.4cm} u,v=t\pm r_*,\cr&&\cr
&&r_*=\int \frac{dr}{f(r)}=\frac{1}{2r_+}\log\left(\frac{r-r_+}{r+r_+}\right),\,\,\,\kappa=r_+
\label{BTZ_Schwarzchild}
\eea
Here $\kappa=2\pi/\beta$ is the temperature of the black hole and is obtained in order to define affine coordinates $\{U,V\}$ at $\rpl$. 
The shockwave metric is quite a simple one if the Kruskal coordinates are known \cite{Dray:1984ha,Sfetsos:1994xa}. One then needs to solve the Einstein's equation 
\be
R_{\mu\nu}-\frac{1}{2}R g_{\mu\nu}+\Lambda g_{\mu\nu}=4\pi G_N T_{\mu\nu},\hspace{0.3cm}T_{\mu\nu}=T_{UU}\sim\delta(U)\delta^{d-2}(\phi-\phi_0) 
\label{Einsteins_eq_shockwave}
\ee
It turns out that the stress-tensor on the right can be absorbed in to the left (i.e. the above eq.  would look like vacuum Einstein's eq.) if we just shifted 
\be
V\rightarrow\tilde{V}=V+h(\phi-\phi_0)\Theta(U)
\label{V_shift_0}
\ee
where $h(\phi)$- a function of transverse coordinates; solves a Laplace(like) equation with a source. The analysis of Shenker \& Stanford \cite{Shenker:2013pqa} assume a spherically symmetric shockwave, therefore the response $h(\phi)\sim 1$ due to spherical symmetry.  
\\\\
In rest of our article, we would be working with the Kruskal extension of the black hole's space-time. The left exterior of the space time is characterized by $U>0$ \& $V<0$ and $vice$-$versa$ for the right with the boundaries at $UV=-1$ and the singularities at $r=0$ implied by $UV=1$. The flow of time on the left boundary is downwards whereas it is upwards on the right boundary with the horizontal space-like surface passing through the bifurcate point defining $t=0$ on both boundaries. We would refer to the shock-wave released from the left boundary at $t>0$ with $U=const$ as being released from the past as compared to the system on the right boundary.
\\\\
Let the shockwave emanate at some time $t_0$ at the boundary $r\rightarrow\infty$. Therefore the metric before and after the shockwave is given by the Kruskal coordinates $\{U,V\}$ and $\{\tilde{U},\tilde{V}\}$ respectively and is of the form  \eqref{BTZ_Schwarzchild}. The shockwave being null would be parametrized by ($r_*\rightarrow 0$ as $r\rightarrow\infty$)
\be
U=-e^{-\kappa t_0},\,\,\,\,\tilde{U}=-e^{-\tilde{\kappa} t_0}\,\,\,\,\&\,\,\,\,
\tilde{r}_+=\sqrt{\frac{M+E}{M}}r_+
\ee
in the two coordinates with $\tilde{\kappa}$ denoting the change in the horizon due to in fallen shockwave and $E$ being the energy of the shockwave at the boundary. One  then demands that the line element along the spherical shockwave is continuous 
\be
r_+\left(\frac{1-UV}{1+UV}\right)=\tilde{r}_+\left(\frac{1-\tilde{U}\tilde{V}}{1+\tilde{U}\tilde{V}}\right)
\ee  
Expanding the above eq. for $E/M\rightarrow 0$ and $t_0\rightarrow\infty$ yields\footnote{One needs to keep $\alpha$ fixed as this limit is taken.}
\be
\tilde{V}=V+\alpha,\,\,\,\,\alpha=\frac{E}{4M}e^{r_+ t_0}
\label{SandS_alpha}
\ee 
Therefore the perturbation grows as $t_0\rightarrow\infty$ with the exponent being $r_+=\kappa$ i.e. the surface gravity or the temperature of the black hole.
\\\\
One then computes the change in the mutual information $I(A:B)$ between equal angular intervals $A$ in CFT$_L$ and $B$ in CFT$_R$. In order to have a non zero unperturbed $I(A:B)$ the angular intervals are less than $\pi$. It is then found that the change in $I(A:B)$ is given by
\bea
I(A:B)&=&S_A+S_B-S_{AB}\cr&&\cr
&=&\frac{l}{G_N}\left[\log \sinh\frac{\pi\phi}{\beta}-\log(1+\alpha)\right]
\label{SandS_I_0}
\eea 
where $S_{AB}$ is sum of the lengths of the 2 geodesics traversing the 2 boundaries of the TFD state in the bulk. The Lyapunov index $\lambda_L$ can be easily read off from the $t_0$ dependence of $I(A:B)$ using \eqref{SandS_alpha} for large black holes
\be
I(A:B)\sim \frac{l}{G_N}\left[\log\sinh\frac{\pi\phi}{\beta}-\kappa t_0+\log\frac{M}{E_0}\right]
\label{SandS_lambda}
\ee
The scrambling time is also determined accordingly to be proportional to $\kappa^{-1}=\beta/2\pi$.
\\\\
It becomes quite clear that the essential feature relevant to  finding the Lyapunov index is the first part of the computation, in that one merely needs to compute the time required for the perturbation in the bulk due to the shockwave to grow exponentially in time. This was also used by Shenker \& Stanford to predict the Lyapunov index in arbitrary higher dim. AdS-Schwarzchild geometries\footnote{Here, in the Appendix A of  \cite{Shenker:2013pqa} the authors only compute the back reaction due to an in-falling shock-wave in Schwarzchild AdS.}.  
\subsection{Rotating BTZ}
The above computation was generalized  for rotating BTZ  by Reynolds and Ross \cite{Reynolds:2016pmi}. The  BTZ metric is 
\be
\frac{ds^2}{l^2}=\frac{dr^2}{f(r)}-f(r)dt^2+r^2\left(d\phi-\frac{r_+r_-}{r^2}dt\right)^2\,\,\,\,w/\,\,f(r)=\frac{(r^2-r_+^2)(r^2-r_-^2)}{r^2}
\label{BTZ_wikipedia}
\ee
The authors of \cite{Reynolds:2016pmi} then choose to redefine
\be
\varphi=\phi-\frac{r_-}{r_+}t
\label{BTZ_corotattion}
\ee 
which co-rotates the boundary with the horizon's angular velocity 
thus yielding
\be
\frac{ds^2}{l^2}=\frac{dr^2}{f(r)}-f(r)dt^2+r^2\left(d\varphi+\frac{r_-(r^2-r_+^2)}{r_+r^2}dt\right)^2
\label{BTZ_counter_rotated}
\ee
Going to Kruskal coordinates implies
\bea
\frac{ds^2}{l^2}
&=&\frac{-4dUdV-4r_-(UdV-VdU)d\varphi+[r_+^2(1-UV)^2+4UV r_-^2]d\varphi^2}{(1+UV)^2}\cr&&\cr
w/&& U=-e^{-\kappa u},\,\,V=e^{\kappa v},\,\,\,\,u,v=t\pm r_*,
\,\,\,\,r_*=
\frac{1}{2\kappa}\log\left(\tfrac{\sqrt{r^2-r_-^2}-\sqrt{r_+^2-r_-^2}}{\sqrt{r^2-r_-^2}+\sqrt{r_+^2-r_-^2}}\right)\cr&&\cr
\& &&\kappa=\frac{r_+^2-r_-^2}{r_+^2}=\frac{2\pi}{\beta}
\label{RandR_kappa}
\eea
where
$\kappa$ is indeed the temperature of the BTZ black hole and repeating the Shenker-Stanford analysis in the above coordinates yields $\lambda_L\leq\kappa=2\pi/\beta$ for cases where analytical solutions are possible to study $I(A:B)$.
\subsection{Possible caveats}
CFT$_2$ has two temperatures $\beta^{-1}_\pm$ (one for each: left moving and right moving)\footnote{$\,\,\,{\rm Note:}\,\,\frac{\beta_+\beta_-}{2\pi}=\beta=\frac{2\pi}{\rpl(1-\mu^2)},\,\,\,\beta_\pm=\frac{2\pi}{\rpl(1\mp\mu)}$ where $\mu=\rmi/\rpl$} and therefore 2 different $\lambda_L$s. This was shown in \cite{Stikonas:2018ane,Poojary:2018eszz,Jahnke:2019gxr} using various methods. Here the authors were primarily concerned with computing out of time ordered correlators (OTOCs) in CFT$_2$. Therefore the above result of \cite{Reynolds:2016pmi} seems to be at odds with these results. Particularly the analysis in \cite{Stikonas:2018ane} does compute the change in $I(A:B)$ $via$ 2 methods: {\bf(1)} In the CFT$_2$ using the prescription of twist operators to compute $I(A:B)$ in a TFD being perturbed by an operator $\psi$ and {\bf (2)} a corresponding bulk analysis where the perturbation due to $\psi$ on rotating BTZ is obtained by mapping the global $AdS_3$ metric perturbed by a massive particle at its origin. This is possible because of the enhanced symmetries of lower dimensional $AdS_3$ $i.e.$ locally $AdS_3$ metrics having no local degrees of freedom. 
\\\\
In the {\bf (1$^{\rm st}$)} approach the author of \cite{Stikonas:2018ane} finds the lower of the 2 temperatures governing the Lyapunov index and the scrambling time, however it becomes apparent from the analysis that the other (higher) temperature could similarly be obtained by choosing a different placement for the perturbation relative to the entangling intervals. We explain this briefly here with reference to the section 2 of \cite{Stikonas:2018ane}. The work in \cite{Stikonas:2018ane} analysis the perturbation due to an operator $\psi$ on the mutual information between two identical  intervals $[y,y+L]$ for $y>0$ in CFT$_L$ and CFT$_R$ in a TFD state  with unequal left (anti-holomorphic) and right (holomorphic) moving temperatures $i.e.$ $\beta_-\neq\beta_+$.  
Here $y$ is the spatial coordinate and the operator $\psi$ is located at $y_w=0$ and time $-t_w<0$. Furthermore, the entanglement entropy is measured for times $t>t_w$. The holomorphic and anti-holomorphic contributions can be analysed separately and are controlled by their respective cross-ratios. The result is analysed as $t$ increases from $t+t_w<y$ to $y<t+t_w<y+L$ to late times $y<t+t_w$. As this happens the anti-holomorphic cross ratio $\tilde{z}$-  sensitive to $\beta_-$, stays close to $1$, while the holomorphic cross ratio $z$- sensitive to $\beta_+$, flips sign thus capturing the effect of the perturbation at late times\footnote{We refer the reader to comments above $eq(2.21)$ in subsection 2.3.1 and above $eq(2.40)$ in subsection 2.3.3 in \cite{Stikonas:2018ane} for the explicit  behaviour of the cross ratios.  }. This is the essential reason for $\beta_+$ appearing in the perturbed mutual information as against $\beta_-$. This can be easily reversed by placing the perturbation  $\psi$ at $y_w>y+L$ instead of $y>y_w=0$ $i.e.$ right of the entangling surface (as depicted in Figure 1. in \cite{Stikonas:2018ane}) as against the  left. One would then find that it is the anti-holomorphic cross-ratio that flips its sign as one increases $t$ thus making the perturbed mutual information depend on $\beta_-$. Therefore the CFT analysis in \cite{Stikonas:2018ane} is consistent with the scrambling time being controlled by 
\be
\lambda_L=\frac{2\pi}{\beta_-}>2\pi T_H.
\ee
This method in the {\bf (2$^{\rm nd}$)} approach is finely tuned to only work in $AdS_3$ unlike the one described in \cite{Shenker:2013pqa} and cannot offer insights as to what may happen in rotating $AdS$ black holes in higher dimensions. As the 2$^{\rm nd}$ method used in   \cite{Stikonas:2018ane} and \cite{Caputa:2015waa} have a one-to-one map with the 1$^{\rm st}$, the change in the placement of the operator $\psi$ must also produce a $\lambda_L>2\pi T_H$ and  subsequently a scrambling time controlled by it. 
\\\\
We would next like to understand the coordinates used in \cite{Reynolds:2016pmi} better as they  seem to lead to a result which is in contradiction with \cite{Stikonas:2018ane}.
As the analysis in \cite{Shenker:2013pqa} relies on shock-waves produced by in-falling null particles, care must be taken to set up Kruskal co-ordinates which trace null in-falling geodesics. It is easily seen that the $\{U,V\}$ coordinates used in \eqref{BTZ_Schwarzchild} are indeed affine coordinates at $\rpl$ and are along in-falling null geodesics with zero angular momenta outside the horizon. However, the $\{U,V\}$ coordinates used in \eqref{RandR_kappa} are not along null geodesics; $i.e.$ $\xi=\partial_U$ (or $\xi=\partial_V$) are not vector fields along null geodesics. These however do furnish good Kruskal coordinates as they are obtained embedding coordinate description of BTZ black holes. This can be seen from the fact that
\be
\xi^\mu\nabla_\mu\xi^\alpha=0 \hspace{0.7cm}{\rm only\,at\,}\,\, r=\rpl.
\ee
Therefore they can be used to extend the coordinates past the horizon onto the the other exterior of the Kruskal extension. However, the above vector fields fail to satisfy the geodesic equations infinitesimally outside the horizon. The analysis of \cite{Shenker:2013pqa} requires one to take the limit of the shockwave solution as the perturbation approaches the horizon. It is precisely this time-scale involved in reaching the outer horizon from the boundary and the subsequent blueshift involved that gives rise to the effect of scrambling of mutual information. A similar analysis for non-rotating BTZ in \cite{Caputa:2015waa} does ensure this. Therefore the null coordinates  in \eqref{RandR_kappa} used in \cite{Reynolds:2016pmi} are not suited for obtaining the Dray-'t Hooft solution. The result of the scrambling time and Lyapunov index as seen in \cite{Reynolds:2016pmi} comes from the exponent used in defining the null coordinates \eqref{RandR_kappa} which as we shall see in the next sub-section can also be obtained by working with null geodesics with zero angular momentum $c.f.$ \eqref{affine_horizon}. Note, that it is not that the coordinates \eqref{RandR_kappa} that are the problem but the fact that the Dray-'t Hooft solution was obtained by working with trajectories which are not null-geodesics  in \cite{Reynolds:2016pmi}. The geodesic equation is coordinate invariant. One could have very well worked with the coordinates \eqref{RandR_kappa} but then the Dray-'t Hooft solution utilising null geodesics would subsequently appear cumbersome.
\\\\
It is also important to note that for rotating geometries one needs to consider null geodesics with non zero angular momentum in order to describe null particles released from the $AdS$ boundary and  falling into the rotating black hole's singularity. In particular we note that in \cite{Cruz:1994ir} it was demonstrated that  for rotating BTZ, at unit energy the null geodesic must have positive angular momenta $\mathcal{L}$ bounded by \footnote{$\frac{2\mu}{1+\mu^2}>\mu\geq 0,\,\forall\,\mu\in [0,1]$ therefore a non-rotating null geodesic in the co-rotating coordinates \eqref{BTZ_corotattion} having $\mathcal{L}=0$ cannot start form the boundary and fall into the black hole. However since we would be working in the limit of late times since the release of the perturbation, we only ought to approximate the trajectory by a null geodesic at late times; that is ideally we need to look at time-like trajectories falling in from the boundary with angular momenta $\mathcal{L}$ which tend to null geodesics at late time as it passes through the horizon. It is reasonable to expect that the lower bound on $\mathcal{L}$ is relaxed.  }
\be
1\geq\mathcal{L}\geq\frac{2\mu}{1+\mu^2},\hspace{0.3cm}\mu=\frac{\rmi}{\rpl} 
\label{L_condition_0}
\ee 
In the next section we set up coordinates along in-falling null geodesics with arbitrary $\mathcal{L}$ and  define  corresponding Kruskal  coordinates $\{U,V\}$ by demanding them to be affine at $\rpl$. We would then set up the Dray-t'Hooft solution in these coordinates for shock-waves with the same value of $\mathcal{L}$ as is required according to the setup described in \cite{Shenker:2013pqa}. 
\section{Rotating Shock-waves}
In this section we first set up coordinates along in-falling null rotating geodesics which are affine at the outer horizon. We then set up the Dray-t'Hooft solution for null rotating shock-waves with the same angular momentum $\mathcal{L}$. We do this for a single null particle and for a thick shell of null particles which is axisymmetric. We also compute the back reaction for the both of them in terms of the time $t_0$ (or $\tau_0=t_0-\mathcal{L}\phi_0$) in the far past of the left boundary. 
\subsection{Affine coordinates}
We would like to compute the blueshift and the back reaction at late times associated to a null rotating shock wave with angular momenta $\mathcal{L}$. We would be interested in those values of $\mathcal{L}$ for which the null geodesic is able to reach the singularity from the boundary of $AdS$. For the case of BTZ geometries this implies \cite{Cruz:1994ir}
\be
1\geq\frac{\mathcal{L}}{\mathcal{E}}>\frac{2\mu}{1+\mu^2},\hspace{0.3cm}\mu=\frac{\rmi}{\rpl} 
\label{L_condition}
\ee 
where $\mathcal{E}$ is the energy of the geodesic.
\\\\
The shockwave computations of Shenker-Stanford for non rotating BTZ and that of \cite{Reynolds:2016pmi} yield $\kappa$ as the $\lambda_L$. This comes about because of the way the Kruskal coordinates are constructed out of $\{t,r_*\}$. The reason why $\{U,V\}$ coordinates are exponentially related to $\{u,v\}$ via the index $\kappa$ has to do with the fact that $\{U,V\}$ are affine coordinates at the horizon.
\\\\
The choice of coordinates depends on the trajectory of the particle which for late enough times is approximated by a null geodesic. Indeed the coordinate system set about a non-rotating null geodesic for a Schwarzchild black hole is the Kruskal coordinate $\{U,V\}$ used above in \cite{Shenker:2013pqa}. The blueshift is then simply determined by defining coordinates $\{U,V\}$ that are affine at the required region, in this case the (near) horizon. Therefore we first need to set up coordinates about rotating null geodesics in rotating BTZ. We first define the vector $\xi^\mu\partial_\mu$ along an arbitrary null geodesic with energy $\mathcal{E}$ and angular momentum $\mathcal{L}$ defined along the killing vectors of the rotating geometry $\zeta_E=\partial_t$ and $\zeta_L=\partial_\phi$
\bea
&&\xi^2=0,\,\,\,g_{\mu\nu}\xi^\mu \zeta_E^\nu=\mathcal{E},\,\,\,g_{\mu\nu}\xi^\mu \zeta_L^\nu=\mathcal{L}\cr&&\cr
\implies&&\xi=\tfrac{1}{r}\sqrt{r^2(\mathcal{E}^2-\mathcal{L}^2)+\mathcal{L}(\mathcal{L}(r_+^2+r_-^2)-2r_+r_-\mathcal{E})}\,\,\partial_r+\frac{\mathcal{E}-\mathcal{N}\mathcal{L}}{f(r)}\partial_t +\cr
&&\hspace{0.8cm}+\,\,\frac{(1-(r_+^2+r_-^2)/r^2)\mathcal{L}-\mathcal{N}\mathcal{E}}{f(r)}\partial_\phi
\label{null_geodesic_0}
\eea  
This also implies that the vector $\xi$ is affine $i.e.$ $\xi\cdot\nabla\xi^\mu=0$. To describe the metric along null vector fields we need to define both in-going and out-going null geodesics. One can be obtained from the other by reversing the direction of the geodesic\footnote{The change in the sign of $\mathcal{E}$ reverses in-falling to out going while change in  the sign of $\mathcal{L}$ is simply due to time reversal.} $i.e.$ $\mathcal{E}\rightarrow-\mathcal{E}$ and $\mathcal{L}\rightarrow-\mathcal{L}$. We denote this pair as $\xi_\pm$ 
 and note that the metric can be expressed in terms of the line elements dual to these vector fields for unit energy ($\mathcal{E}=1$)
\bea
ds^2_{BTZ}&=&F(r)\,\,\xi^+_\mu dx^\mu\,\xi^-_\nu dx^\nu + h(r)\left(d\phi+\tilde{h}_1(r)d\tau\right)^2\cr&&\cr
&=& F(r) \,du\,dv + h(r)\left(d\phi+\tilde{h}_1(r)d\tau\right)^2\cr&&\cr
&{\rm where}&\,\,\xi^\pm\cdot dx=dr_*\pm d\tau,\,\,\,{\rm \&}\,\,u=r_*-\tau,\,\,\,\,v=r_*+\tau,\,\,\,\, \tau=t-\mathcal{L}\,\phi,\cr&&\cr
&&r_*=\int_r^\infty\frac{dr}{rf(r)}\sqrt{r^2(1-\mathcal{L}^2)+\mathcal{L}(\mathcal{L}(r_+^2+r_-^2)-2r_+r_-)},\cr&&\cr
&&F(r)=\frac{(r^2-\rpl^2)(r^2-\rpl^2\mu^2)}{r^2(1-\mathcal{L}^2)+\rpl^2(\mathcal{L}(1+\mu^2)-2\mu)}
\label{LC_BTZ}
\eea
Here $r_*$ is the relevant tortoise coordinate for such a coordinate along rotating null geodesics.
As can be checked the light-cone coordinates $\{u,v\}$ are not affine  $i.e.$  $\chi_u\cdot\nabla\chi_u^\mu=\mathcal{K}\chi_u^\mu$ for $\chi_u=\partial_u,\,\,$ similarly for $\chi_v=\partial_v$. \emph{Note}: $\partial_u$ is not the same as $\xi_+$ but is indeed proportional to $\xi_+$ $i.e.$ $\chi_u=F\xi_+$. This in turn implies 
\be
\mathcal{K}=\left|\tfrac{1}{2}\xi_\pm\cdot\partial F\right|
\label{kappa}
\ee
We can now define affine coordinates $\{U,V\}$ at the horizon by
\be
\boxed{
U=-e^{\kappa \,u},\,\,\,\,V=e^{\kappa\, v},\,\,{\rm where}\,\,\,\kappa=\mathcal{K}\vert_{r_+}=\frac{\rpl(1-\mu^2)}{(1-\mu\mathcal{L})}=\frac{2\pi}{\beta(1-\mu\mathcal{L})}
}
\label{affine_horizon}
\ee
The above coordinates capture the right exterior defined by $U<0<V$. The left exterior $U>0>V$ is obtained by reversing the signs of $\{U,V\}$. 
This is the analogue of Kruskal coordinates as seen by a null particle falling into the black hole with an angular momentum $\mathcal{L}$. We see that for non-rotating BTZ ($\mu=0$), $\kappa$ is the black hole's temperature for any value of $\mathcal{L}$.
The metric can then be written in terms of the above coordinates as
\bea
ds^2&=&\frac{F}{\kappa^2 UV} \,dUdV +h[dz+\frac{h_1}{2\kappa UV}\left(UdV-VdU)\right]^2,\cr&&\cr
{\rm with}\,\,\, z&=&\phi-\mu t,\,\,\,\,\,\tau =t-\mathcal{L}\phi 
\label{affine_coord}
\eea
where the functions $F,h$ \& $h_1$ depend only on $r$ $via$ $UV$. The choice of $z$ is forced by demanding that $h_1/UV$ be finite as $U\rightarrow 0$ (or $V\rightarrow 0$). The instantaneous angular coordinate of  the particle $z_p$ falling along $U=U_0$ is given by vanishing of the the transverse direction in \eqref{affine_coord}
\bea
&&dz_p+h_1(U_0 V)\frac{dV}{2\kappa V}=0\cr&&\cr
{\rm as}\,&& h_1\underset{U_0\rightarrow 0}{\longrightarrow}0\implies dz_p\underset{U_0\rightarrow 0}{\longrightarrow}0
\label{phi_p}
\eea 
We note its value as $U_0\rightarrow 0$ for later purposes. The fact $h_1\underset{U_0\rightarrow 0}{\longrightarrow}0$ can be seen as a consequence of having the transverse line element finite at the horizon ($c.f.$ \eqref{r_UV}). 
\\\\
We mention the blueshift along rotating geodesics for RN-AdS$_4$ and Kerr-AdS$_4$ in appendices \eqref{RN} and \eqref{Kerr} respectively as preliminary results  of ongoing work \cite{Malvimat:2022ongoing}.
\subsubsection{Tortoise close to the horizon}
It would be useful that we have a relation between the Boyer-Lindquist and tortoise coordinates close to the horizon for later purposes. We note that the relation \eqref{kappa} using \eqref{LC_BTZ} can be written as
\bea
&&\mathcal{K}=\tfrac{1}{2}g^{r_* r_*}\xi^\pm_{r_*}\partial_{r_*}F,\hspace{0.7cm}
\implies \left.\partial_{r_*}\log F\right\vert_{\rpl}=2\kappa.
\eea  
Expanding $F$ to linear order at the horizon $r_+$ we have\footnote{\emph{Note} that for $\frac{F}{\kappa^2UV}dUdV$ in \eqref{affine_coord} to be finite at $U\rightarrow 0$, $F|_{\rpl}$ has a zero of $\mathcal{O}(r-\rpl)$ as can be seen from \eqref{LC_BTZ}. It is the coefficient of this zero that determines the relation between $r$ and $UV$ in the near horizon.}
\be
F(r)\sim(r-r_+)F'(r_+)=-4\kappa^2e^{2 \kappa\, r_*}=-4 \kappa^2 U V.
\label{r_UV}
\ee
This allows us to express $(r-\rpl)$ in terms of the combination $U V$. This relation can be further used in 
\be
F(r)\sim(r-r_+)F'(r_+)+\tfrac{1}{2}(r-r_+)^2F''(r_+)
\ee
to obtain the value of 
\be
\left.\partial_{UV}\left(\frac{F(r)}{\kappa^2 UV}\right)\right\vert_{\rpl}=\frac{16\kappa^2}{F'(\rpl)^2}F''(\rpl)
\ee
which would be used later for computing the back-reaction to a shock wave.
\subsubsection{Blueshift}
An intuitive argument to understand the result of \cite{Shenker:2013pqa} is to compute the blueshift seen by an in-falling null particle as a function of the time $t_0$ at which it was released into the bulk with infinitesimally small energy $E_0$. 
\be
E\sim E_0 e^{\frac{2\pi}{\beta}t_0}
\label{SandS_blueshift}
\ee
The Lyapunov index and the time taken to disrupt a non-zero mutual information between the 2 CFTs in a TFD state is then found to be governed by this time dependence. We try to understand this intuition by  computing the blue shift suffered by a null particle falling into the black hole with arbitrary angular momentum $\mathcal{L}$.  
\\\\  
For an in-falling particle along the affine co-ordinate $V$ the trajectory is defined by $U=U_0$ and initial angular coordinate $\phi_p\vert_{r=\infty}=\phi_0$. This is most easily obtained by noting that the light-cone coordinates close to the horizon $\{u,v\}$ and that on the boundary are related by a Rindler transformation \eqref{affine_horizon}. For this we only need to analyse the light-cone directions of the metric and note that close to the boundary the metric in the $\{u,v\}$ directions is of the form
\be
ds^2=\frac{1}{\epsilon^2}\,dudv
\ee 
where $\epsilon\rightarrow 0$ defines the boundary. The same is true of the metric written interms of the affine coordinates at the horizon $i.e.$
\be
ds^2=\, dUdV
\ee   
while the metric at the horizon written in terms of $\{u,v\}$ vanishes as $F(r_+)=0$. Therefore the affine coordinates at the horizon represent the Minkowski coordinates similar to those at the boundary. In other words we ought to work with those set of coordinates which are smooth at the horizon.  The blueshift at the horizon can be hence be read off from the Rindler transformation \eqref{affine_horizon} relating the $\{U,V\}$  to the $\{u,v\}$ coordinates. If the null momentum of the particle in the affine frame is $E_0k^V$ then we have 
\be
E_pk^v=\frac{E_0}{\kappa V}\, k^V,
\label{boost_Ep}
\ee
thus the boosted energy  $E_p$ at the horizon along $U=U_0$ is
\be
E_p=\left.\frac{E_0}{\kappa V}\right\vert_{U_0}=\frac{E_0e^{-2\kappa\tau}}{U_0}\overset{\tau=0}{=}E_0\,e^{\kappa \tau_0},\,\,\,\,{\rm with}\,\,\,\kappa=\frac{2\pi}{\beta(1-\mu\mathcal{L})}
\label{boost_U0}
\ee
This has the the expected $1/U_0=e^{\kappa\, \tau_0}$ behaviour for the blueshift of the particle as it passes the $t=0$ slice when thrown in at a time $t_0=\tau_0+\mathcal{L}\phi_0$ in the past from the right boundary.  The above result smoothly reproduces the results expected for non-rotating black holes and non-rotating null geodesics. 
\\\\
It is important to note that this value of $\kappa$ is not simply an artefact of having to work with $\{\tau,z\}$ instead of $\{t,\phi\}$, it is physical as can be seen from the blueshift above and survives the extremal limit for $\mu\rightarrow 1,\mathcal{L}\rightarrow 1$, in which case $\kappa=2\rpl$.
\\\\
Given the above behaviour of blueshift along null in-falling geodesics with angular momentum $\mathcal{L}$ we expect- like in the analysis in \cite{Shenker:2013pqa}, that the back-reacted metric  must also capture the dependence on $\kappa$  at late times. The shockwave solution at late times is obtained by using the Dray -'t Hooft solution as seen next. 
\subsection{Shockwave solution}
We next write down a shockwave solution for the metric which suffers a back reaction owing to an in-falling null particle and a thick shell of null particles released in the far past. For this 
we first  note the Dray -'t Hooft solution for a shockwave along the in-falling affine coordinate $V$ at $U=0$ is given by
\bea
&&ds^2=\frac{F}{\kappa^2UV} \,dUdV +h(dz+\frac{h_1}{2\kappa U V}\left(UdV-VdU)\right)^2 
\,\,\rightarrow ds^2+\alpha \frac{F}{\kappa^2 U\tilde{V}}\delta(U)f(z)dU^2\cr&&\cr
{\rm where}\,&&V\rightarrow\tilde{V}=V+\alpha f(z)\Theta(U),\,\,\,\left.\partial_V \tfrac{F}{UV}\right\vert_{U=0}=0,\,\,\left.\partial_V h\right\vert_{U=0}=0.
\label{shockwave_0}
\eea 
Here $\alpha$ is the energy of the shockwave measured locally. Here the transverse line element is not effected by the diffeomorphism $V\rightarrow\tilde{V}$ as $h_1$ vanishes at $U=0$ (due to our choice of $z$ $c.f.$ below \eqref{affine_coord}.).
\subsubsection{Single particle}
We first write an anzats for the  metric with a null particle at $U=U_0\neq 0$.
\bea
&&ds^2=\frac{F}{\kappa^2UV} \,dUdV +h(dz+\frac{h_1}{2\kappa U V}\left(UdV-VdU)\right)^2 
+\alpha \frac{F}{\kappa^2 U\tilde{V}}\delta(U-U_0)f(z)dU^2\cr&&\cr
{\rm where}\,&&V\rightarrow\tilde{V}=V+\alpha f(z)\Theta(U-U_0),
\label{shockwave_U0}
\eea 
where $U_0=e^{-\kappa \tau_0}$ and $\tau_0=t_0-\mathcal{L}\phi_0$ defines the time and space co-ordinate on the left boundary at which the the null particle is released. 
We expect the above metric to match the metric \eqref{shockwave_0} as we take $U_0\rightarrow 0$ $i.e.$ as $\tau_0\rightarrow\infty$. This would correspond to sending in the shockwave at $t>>0$ from the left boundary which corresponds to the far past $w.r.t.$ the right exterior. We determine $\alpha$ be demanding continuity in the transverse volume along the shockwave
\be
\left. h(U V)\right\vert_{U_0^-}=\left. h(U \tilde{V})\right\vert_{U_0^+}
\label{single_smoothness}
\ee 
We simplify the above condition in the limit 
\bea
&&U_0\rightarrow 0,\,\,\,\, \frac{\delta r_+}{r_+}\rightarrow 0,\cr&&\cr
{\rm with },\,\,\,&&\frac{\delta r_+}{U_0 r_+}\sim finite
\label{limit_horizon_change_U0}
\eea
where $\delta r_+$ is the infinitesimal change in the outer horizon $r_+$. We expect this change to be small and is equivalent to demanding $E_0/M\sim 0$, $E_0$ being the energy of the shockwave as measured at the boundary and $M$ being the mass of the black hole. We thus get a relation between $V$ and $\tilde{V}$ in this limit to be
\be
\tilde{V}=V+\frac{\delta r_+}{U_0r_+}\left(\tfrac{r_+ F'(r_+)}{\kappa^2}\right)
\label{V_shift}
\ee  
comparing with the shift in $V$ in \eqref{shockwave_0} we note 
\be
\alpha f(z)\sim \frac{\delta r_+}{U_0 r_+}
\ee
It is worth pausing and noting the following facts: {\bf(a)} The shift in the outer horizon at late times only depends on $z$ $via$ $f(z)$ {\bf(b)} The shift in the $V$ coordinate along the shockwave at late times is independent of $V$ and goes as $1/U_0$. The precise nature of the relation between $E_0$ and $\delta r_+$ although interesting would not effect the physics that concerns  us. We therefore write $\alpha$ as
\be
\alpha=\frac{E_{eff}}{U_0}
\ee 
where $E_{eff}=|\tfrac{\delta r_+}{r_+}|\sim \tfrac{E_0}{M}$ with $||$ being the average over $z$. The limit \eqref{limit_horizon_change_U0} is then equivalent to
\bea
U_0\rightarrow 0,\,\,\,\, E_{eff}\rightarrow 0,\,\,\,\,{\rm with },\,\,\,&&\frac{E_{eff}}{U_0 }\sim finite.
\label{limit_horizon_change_U0}
\eea
Einstein's equation can then be solved for backreaction $f(z)$ in the transverse direction
\be
\alpha\,\delta(U-U_0)\,\mathcal{D}f(z)=4\pi G_N T_{UU}=\frac{E_{eff}}{U_0}\,\delta(U-U_0)\delta(z-z_p)
\label{eom_backreaction}
\ee
where we absorb $G_N$ in the definition of $E_{eff}$ as it defines the energy scale with which we perturbed the system at late times. The above equation constrains the backreaction in the $z$ direction given by the instantaneous  location $z_p$ of the null particle. 
\\\\
The position of the null particle  is governed by vanishing of the transverse line element $dy=[dz+h_1d\tau]$.
At late times ($U_0\rightarrow 0$) this implies \eqref{phi_p} $i.e.$
\be
y_p=z_p=const
\ee
Evaluating the line element $dy$ in terms of the $\{dt,d\phi\}$ for $r\rightarrow \infty,\,\, t_{const.}$ implies
\be
\left.dy_p\right\vert_{r\rightarrow\infty,t}=\left(1-(1-\mu\mathcal{L})h_1(1)\mathcal{L}\right)d\phi
\ee
As the transverse position of the shockwave is always given by $dy_p=0$ we have
\be
z_p=\tfrac{1}{1-\mathcal{L}^2}\,\phi_0=z_0
\ee
where $\phi_0$ is the position of the shockwave as it starts out from the boundary and $h_1(1)$ is the value of $h_1$ at the boundary. Therefore we have
\be
\mathcal{D}f(z)=\delta(z-z_0)
\label{back_reaction_f}
\ee
wherein crucially there is no dependence on the time $t_0\sim \log\,U_0$ at which the shockwave was sent in from the boundary. We analyse the solution for the single particle back reaction in the appendix.
\subsubsection{Thick-Thin Shell} We next analyse the back reaction due to null particles released from every point on the boundary simultaneously at time $t_0$ from the left boundary with angular momentum $\mathcal{L}$. This would imply a thick shell of null particles defined by $\tau_{sh}=[t_0,t_0-2\pi\mathcal{L}]$, in terms of the Kruskal coordinates we have
\be
U_{sh}=[U_0,U_0e^{2\pi\kappa\mathcal{L}}], \,\,\,\,U_0=e^{-\kappa t_0}
\label{U_shell}
\ee 
However we would be interested in releasing the particles in the far past $t_0\rightarrow\infty$, thus as $U_0\rightarrow 0$ the thick shell in $U$ at $U_{sh}$ becomes a thin shell at $U=0$. Therefore at late times we expect the metric to take the form \eqref{shockwave_0} with $f(z)=const.\sim 1$ as $z$ is the co-moving periodic coordinate and the shockwave is present at every value of $z$. For intermediate times we write down the metric due to the thick shell as
\bea
ds^2&=&\frac{F}{\kappa^2UV} \,dUdV +h(dz+\frac{h_1}{2\kappa U V}\left(UdV-VdU)\right)^2 +\cr&&\cr
&&\hspace{0.1cm}+\alpha \frac{F}{\kappa^2 UV}\left[\frac{\Theta(U-U_0e^{2\pi\mathcal{L}\kappa})-\Theta(U-U_0)}{U_0(e^{2\pi\mathcal{L}\kappa}-1)}\right]dU^2.
\label{shockwave_thick}
\eea  
where we can see that the quantity in the box brackets above tends to $\delta(U)$ as $U_0\rightarrow 0$, in which case the metric after the shock wave becomes
\bea
ds^2\rightarrow ds^2&=&\frac{F}{\kappa^2UV} \,dUdV +h(dz+\frac{h_1}{2\kappa U V}\left(UdV-VdU)\right)^2 
+\alpha \frac{F}{\kappa^2 U\tilde{V}}\delta(U)dU^2\cr&&\cr
{\rm where}\,&&V\rightarrow\tilde{V}=V+\alpha \Theta(U)
\label{shockwave_thin}
\eea
The smoothness of the transverse direction can similarly be imposed across the thick shell to obtain the $U_0$ dependence of $\alpha$. Like \eqref{single_smoothness} this implies
\be
\left. h(UV)\right\vert_{U_0^-}+ \left. \Delta h(UV)\right\vert_{U_0^-}=\left. h(U\tilde{V})\right\vert_{U_1^+}
\label{smoothness_thick}
\ee
where $U_1=U_0e^{2\pi\mathcal{L}\kappa}$ and $\left.\Delta h(UV)\right\vert_{U_0^-}$ is the change in $h(U_0^-V)$ due to change in $U_{sh}$ across the shell. Simplifying the above matching condition in the limit \eqref{limit_horizon_change_U0} we find
\be
\tilde{V}=V+\frac{\delta r_+}{r_+U_0\,e^{2\pi\kappa\mathcal{L}}}\left(\tfrac{r_+ F'(r_+)}{\kappa^2}\right)
\label{V_shift_thick}
\ee
Comparing with \eqref{shockwave_thin} we find
\be
\alpha\sim \frac{\delta r_+}{r_+U_0\,e^{2\pi\kappa\mathcal{L}}}\sim\frac{E_{eff}}{U_0}
\ee
Here, like in the single particle case $\delta r_+$ is the change in outer horizon due to the collapsed shell and would be proportional to the total energy $E_0$ with which the shell started out from the boundary. \\\\
The inverse dependence of $\alpha$ on $U_0$ is the same as expected form the blueshift of an in-falling null particle at late times. It is this dependence on $\kappa$ $via$ $U_0=e^{-\kappa \tau_0 }$ that captures the required dependence of the  back reacted metric. However like in \cite{Shenker:2013pqa} this change is only perceptible if one probes lengths across the future horizon from the right exterior. Therefore RT-surfaces measuring  Mutual Information between large enough subsystems in the left and right dual CFTs should be sensitive to this $\kappa\geq\frac{2\pi}{\beta}$.        

\section{Mutual Information}

As described earlier we are going to compute the effect of the shock wave on the holographic mutual information between two subregions $A$ and $B$, one on each boundary $CFT$ i.e let us say $A$ is in $CFT_L$ and $B$ is in $CFT_R$. The mutual information is an algebraic sum of entanglement entropies $S_A,S_B$ and $S_{AB}$ given as follows
\begin{align}
I(A:B)=S_A+S_B-S_{AB}
\end{align}
In the context of $AdS_3/CFT_2$, the subregions are angular intervals and each of the entanglement entropies are given by the length of geodesic/geodesics homologous to the corresponding interval. We will choose the intervals $A$ and $B$ of equal length and take them large enough such that the entanglement wedge of the subsystem $AB$ is connected ( if the entanglement wedge is disconnected $I(A:B)=0$ ) and the geodesics corresponding to $S_{AB}$ traverse from one boundary $CFT$ to another. As these surfaces cross the future horizon of the right exterior they will notice the abrupt coordinate change at $U=0$ in $V$.   Note that at late times the RT surfaces/ geodesics corresponding to the subregions $A$ and $B$ are unaffected by the presence of shockwaves  and are given by
\begin{align}
S_A=S_{B}&=\frac{\gamma_{AB}}{4 G_N}=\frac{l}{4G_N}\log \left[\frac{4 r_0^2}{(r_+^2-r_-^2) } \sinh (\frac{(r_++r_-) \phi_L}{2}) \sinh (\frac{(r_+-r_- )\phi_L}{2})\right],\\
&=\frac{c}{6} \log \left[\frac{\beta_{-} \beta_{+}}{\pi^{2} \epsilon^{2}} \sinh( \frac{\pi \phi_L}{\beta_{-}}) \sinh (\frac{\pi \phi_L}{\beta_{+}})\right]
\label{S_A}
\end{align}
 where $\phi_L$ is the angular length of the intervals  $A$ and $B$ which we have chosen to be same and $\beta_+,\beta_-$ are the left and right moving temperatures of the boundary $CFT$ and are related to $r_+$ and $r_-$ as $\beta_{\pm}=\frac{2\pi}{ r_+ \pm \,r_-}$. Note that in order to arrive from the first line to second we have utilized the  Brown-Henneaux \cite{Brown:1986nw} formula $c=\frac{3 l }{2 G_N}$ and the UV-IR relation $r_0\sim \frac{1}{\epsilon}$ where $r_0$ is bulk infrared cut-off and $\epsilon$ is the UV cut-off of the boundary $CFT$.
\\\\
We would also find it convenient to compute the mutual information in terms of the co-moving frame defined by $z=\phi-\mu t$ and $\tau=t-\mathcal{L}\phi$ at the boundary. Since we would be interested in computing    $I(A:B)$ at a fixed time $t=0$ at both boundaries  we have $\delta z_{L,R}=\delta \phi_{L,R}$. This is crucial as it is the $z$ coordinate that is left unaffected by the shock-wave at late times. Therefore the expression \eqref{S_A} holds true in the presence of a shock-wave.

\subsection{Computation of $S_{AB}$}

  Let us now focus on the geodesic that traverses from one boundary $CFT$ to another through the bulk black hole space time.
Since $AdS_3$ enjoys symmetries of a group manifold, any BTZ geometry can be obtained from the embedding coordinates in $\mathbb{R}^{2,2}$ describing a time-like hyperbola $-T_{-1}^2-T_0^2+X_1^2+X^2_2=-l^2$ with $l$ being the radius of $AdS_3$. Any BTZ metric is obtained by the $\mathbb{R}^{2,2}$ metric pulled back on this surface. The embedding coordinates can be parametrized in terms of BTZ coordinates as 
\bea
T_0= \sqrt{\frac{r^2-r_+^2}{r^2_+-r_-^2}}\sinh(r_+ t-r_-\phi),&&\hspace{0.2cm}T_{-1}= \sqrt{\frac{r^2-r_-^2}{r^2_+-r_-^2}}\cosh(r_+ \phi-r_-t)\cr&&\cr
X_1= \sqrt{\frac{r^2-r_+^2}{r^2_+-r_-^2}}\cosh(r_+ t-r_-\phi),&&\hspace{0.2cm}X_2= \sqrt{\frac{r^2-r_-^2}{r^2_+-r_-^2}}\sinh(r_+ \phi-r_-t).
\eea
For any point close to the horizon the above embedding coordinates can be expressed in terms of $\{U,V,x\}$, where we use \eqref{affine_coord}\eqref{affine_horizon}\eqref{r_UV}. At the horizon $U=0$ we have
\bea
T_0=X_1=\sqrt{\tfrac{2r_+}{r_+^2-r_-^2}}\,\frac{\kappa\,\, e^{-r_+\frac{(\mu-\mathcal{L})}{1-\mu\mathcal{L}}z}}{F'(r_+)}V,\hspace{0.2cm}T_{-1}=\cosh(z),\hspace{0.2cm}X_2=\sinh(z)
\eea
 We first find the geodesic distance from a point $p_{\partial_R}=(r,t_R,\phi_R)\simeq(r,\tau_R,z_R)$ on the boundary at $(r\rightarrow\infty)$ to a point $p_{r_+}=(0,V,z)$ on the horizon $(U=0)$.  This is
\be
\cosh (\tfrac{d_{(p_{r_+},p_{\partial_R})}}{l})\sim\frac{r}{\sqrt{r_+^2-r_-^2}}\left[\cosh(\rpl(z_R- z))-\sqrt{\tfrac{2r_+}{r_+^2-r_-^2}}\frac{\kappa\,\, e^{\rpl\tfrac{\mu-\mathcal{L}}{1-\mu\mathcal{L}}(z_R- z)} e^{-\kappa\tau_L}}{\sqrt{F'(r_+)}}V\right]
\label{right_geodesic}
\ee 
where we have kept only the divergent contribution as  $p_{\partial_R}\rightarrow\partial_R$. Similarly, the geodesic from the left boundary point $p_{\partial_L}=(r\rightarrow\infty,t_R,\phi_R)$ can be obtained by changing the signs of $T_0$ \& $X_1$. This is
\be
\cosh (\tfrac{d_{(p_{r_+},p_{\partial_L})}}{l})\sim\frac{r}{\sqrt{r_+^2-r_-^2}}\left[\cosh(\rpl(z_L- z))+\sqrt{\tfrac{2r_+}{r_+^2-r_-^2}}\frac{\kappa\,\, e^{\rpl\tfrac{\mu-\mathcal{L}}{1-\mu\mathcal{L}}(z_L- z)}e^{-\kappa\tau_R} }{\sqrt{F'(r_+)}}(V+\alpha f(z))\right]
\label{left_geodesic}
\ee 
where we have shifted the $V$ coordinate as this region falls in the past of the light-cone of the null particle. For convenience we take $z_R=z_L$ and $\tau_L=\tau_R$. Extremizing with respect to $V$ yields $V=-\tfrac{1}{2}\alpha f(z)$. This implies that the two lengths are equal, and for large values of $r$  for the boundary points, it is given by
\bea
\frac{d_L+d_R}{l}&=&2\log\left[2\frac{r}{\sqrt{r_+^2-r_-^2}}\right]+
\cr&&\cr
&&+2\log\left[ \cosh(\rpl(z_L-z))+e^{\rpl\tfrac{\mu-\mathcal{L}}{1-\mu\mathcal{L}}(z_L-z)}\, \frac{\alpha e^{-\kappa\tau_L}}{2} f(z)\right]
\label{length_x_unextremized}
\eea 
where we have yet to extremize the intermediate point $z$ at the horizon. Here the second line captures the effect of the shock wave. We next need to minimize  
\be
\cosh(\rpl(z_L-z))+e^{\rpl\tfrac{\mu-\mathcal{L}}{1-\mu\mathcal{L}}(z_L-z)}\, \frac{\alpha e^{-\kappa\tau_L}}{2} f(z)
\label{func_minima}
\ee
for $f(z)$ given in \eqref{f_sol} for a single particle and $f(z)=1$ for the thick-thin shell. Here we have assumed without loss of generality that the shock wave due to a single particle is present at $z_0$ at $t=0$. We also need to impose the condition  \eqref{L_condition} on $\mathcal{L}$ for the shockwave as we would be interested in only those null trajectories that can fall into the black hole from the boundary. Therefore we choose an $\mathcal{L}$ constrained by  
\be
1\geq\mathcal{L}\geq\frac{2\mu}{1+\mu^2}
\label{L_condition_1}
\ee
and then extremize $w.r.t$ $z$. 

\subsubsection{$\mathcal{L}=1$ Shell} Analytic solutions can be easily found for the  shell with $\mathcal{L}=1$. Eq. \!\!\eqref{func_minima} reduces to extremizing the following expression
\be
\cosh(\rpl(z_L-z))+\frac{\alpha e^{-\kappa \tau_L}}{2}e^{-\rpl(z_L-z)}
\ee
which implies $e^{\rpl(z-z_L)}=(1+\alpha e^{-\kappa\tau_L})^{-1/2}$. The length of the traversing geodesic at $t_L=0(\implies \tau_L=\kappa\mathcal{L}z_L)$ is then given by 
\bea
\implies&&
\frac{d_{L+R}}{l}=2\log
\left[\frac{2r}{\sqrt{\rpl^2-\rmi^2}}\right]+2\log\left[\sqrt{1+\alpha e^{\kappa \mathcal{L}z_L}}\right]
\eea
\subsection{Mutual Information}
We first analyse the change in $I(A:B)$ due to the shock-wave shell with $\mathcal{L}=1$.
The mutual information for large enough intervals is therefore
\bea
I(A:B)_{\mathcal{L}=1}&=&S_A+S_B-S_{AB}\cr&&\cr
&=&\frac{l}{2G_N}\log\left[\sinh(\frac{\pi \delta\phi}{\beta_-})\sinh(\frac{\pi \delta\phi}{\beta_+})\right]-\frac{l}{4G_N}\log\left[(1+\alpha e^{\kappa\mathcal{L} z_{L_2}})(1+\alpha e^{\kappa\mathcal{L} z_{L_2}})\right]\cr&&\cr
{\rm with}&& \alpha=\frac{E_{eff}}{M}e^{\kappa\, t_0},\hspace{0.3cm}\delta\phi=(\phi_{L_1}-\phi_{L_2})=z_{L_1}-z_{L_2} \,\,{\rm at}\,\, t=0
\label{I_shell_L1}
\eea
We would like to compare this with the case originally studied by Shenker and Stanford \cite{Shenker:2013pqa}  with $\mu=0=\mathcal{L}(\implies \kappa=2\pi/\beta)$ where  mutual information is given by
\bea
I(A:B)_{\mu=0=\mathcal{L}}&=&\frac{l}{G_N}\left\lbrace\log\sinh\left(\frac{\pi\delta\phi_L}{\beta}\right)-\log[1+\alpha]\right\rbrace\cr&&\cr
{\rm with}&& \alpha=\frac{E_0}{M}e^{\kappa\, t_0}, 
\label{Lyapunov_shell_L1}
\eea
The Lyapunov exponent can be simply read off from the  $t_0$ dependence of $I(A:B)_{\mu=0=\mathcal{L}}$ for large black holes
\be
I(A:B)_{\mu=0=\mathcal{L}}\sim S_A +S_B- \frac{l\kappa}{G_N}t_0 -\log \frac{E_0}{M}\implies \lambda_L=\kappa=\frac{2\pi}{\beta}
\label{Lyapunov_SandS}
\ee  
The scrambling time is also determined by $\kappa=2\pi/\beta$ in this case. We see from \eqref{I_shell_L1} that
\bea
I(A:B)_{\mathcal{L}=1}&\sim &S_A +S_B-\frac{l\kappa}{2G_N}t_0-\log\frac{E_{eff}e^{\kappa\mathcal{L}(z_{L1}+z_{L_2})}}{M}\cr&&\cr
\implies && \lambda_L=\frac{\kappa}{2}=\frac{\pi}{\beta(1-\mu)}
\eea
Taking $z_{L_1}+z_{L_2}=0$ $i.e.$ an interval equidistant from $z=0$ and writing $M\sim S/\beta$ we have
\be
t_*=\frac{\beta(1-\mu)}{\pi}\log \frac{S}{\beta E_{eff}}
\ee
The scrambling time in this case is therefore determined by $\kappa/2$ as $E_{eff}\sim E_0 $ can be taken to be of the order of black hole's temperature. We particularly note that for $\mu>1/2$ this is greater than the temperature of the black hole $i.e.$
\be
\mathcal{L}=1,\mu>\frac{1}{2}\implies\frac{\kappa}{2}> \frac{2\pi}{\beta}
\label{shell_kappa_analytic}
\ee
Given the analysis in the previous section of how the blueshift is determined by $\kappa$ one would expect  $\lambda_L$ to be bounded by $\kappa$ and not $\kappa/2$. We can see this in the specific example of $\mu=0$ $i.e.$ Schwarzchild black hole that $\lambda_L|_{\mathcal{L}=0}=\kappa$ while $\lambda_L|_{\mathcal{L}=1}=\kappa/2$ \eqref{Lyapunov_shell_L1}. Therefore one may suspect that the Lyapunov index being half the blueshift at the horizon may be related to peculiarities of the geometry of the subsystem and $\mathcal{L}=1$. 
\begin{figure}
\includegraphics[scale=0.25]{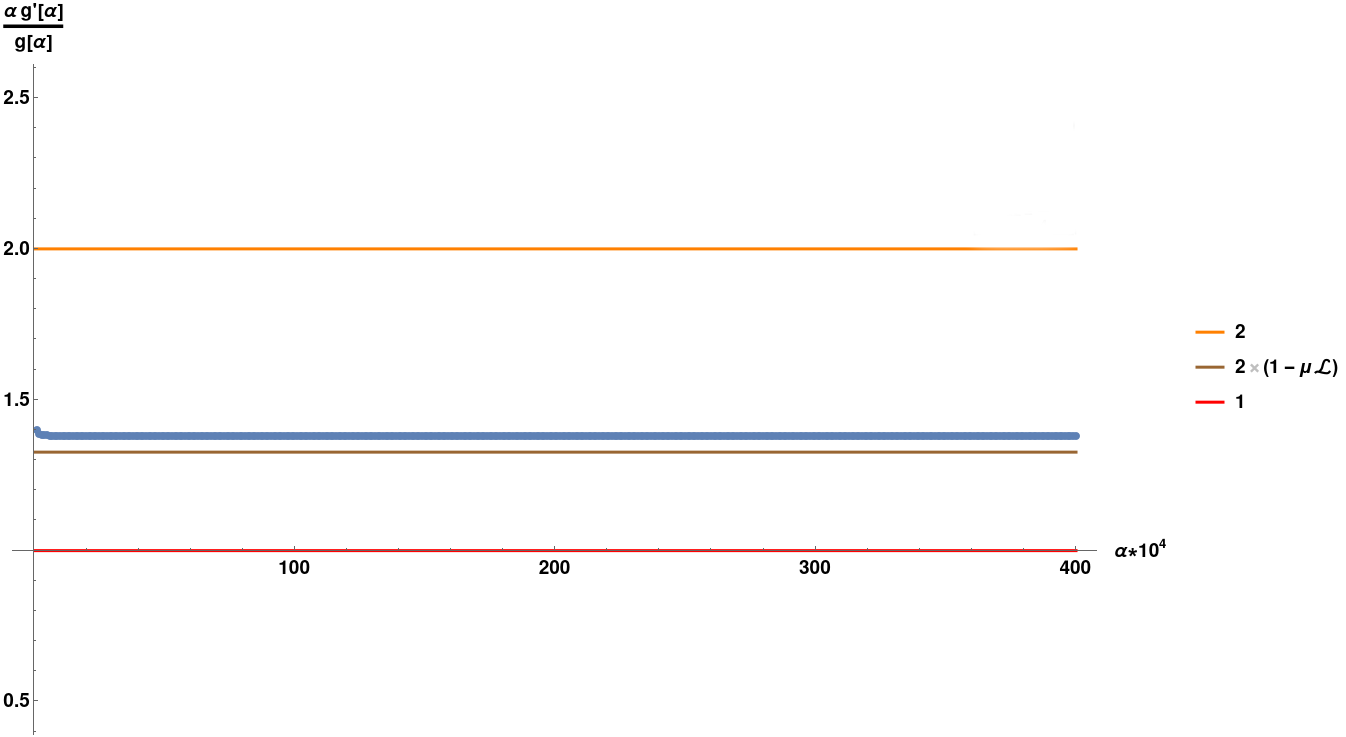}
\includegraphics[scale=0.25]{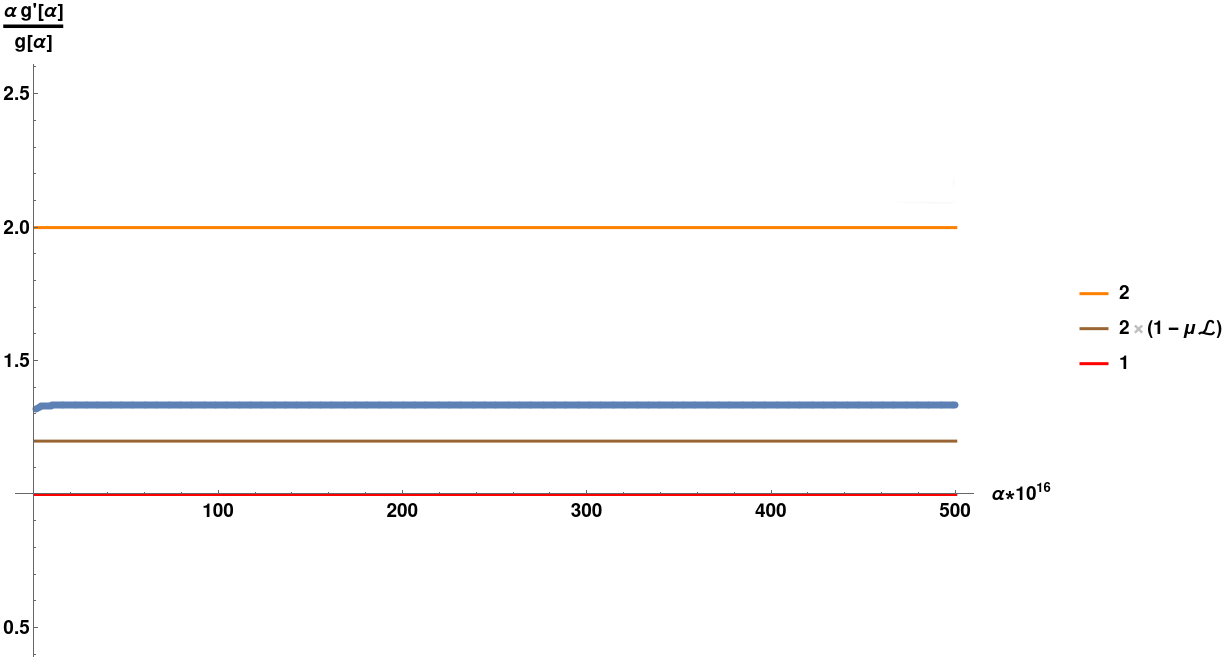}\\
\includegraphics[scale=0.25]{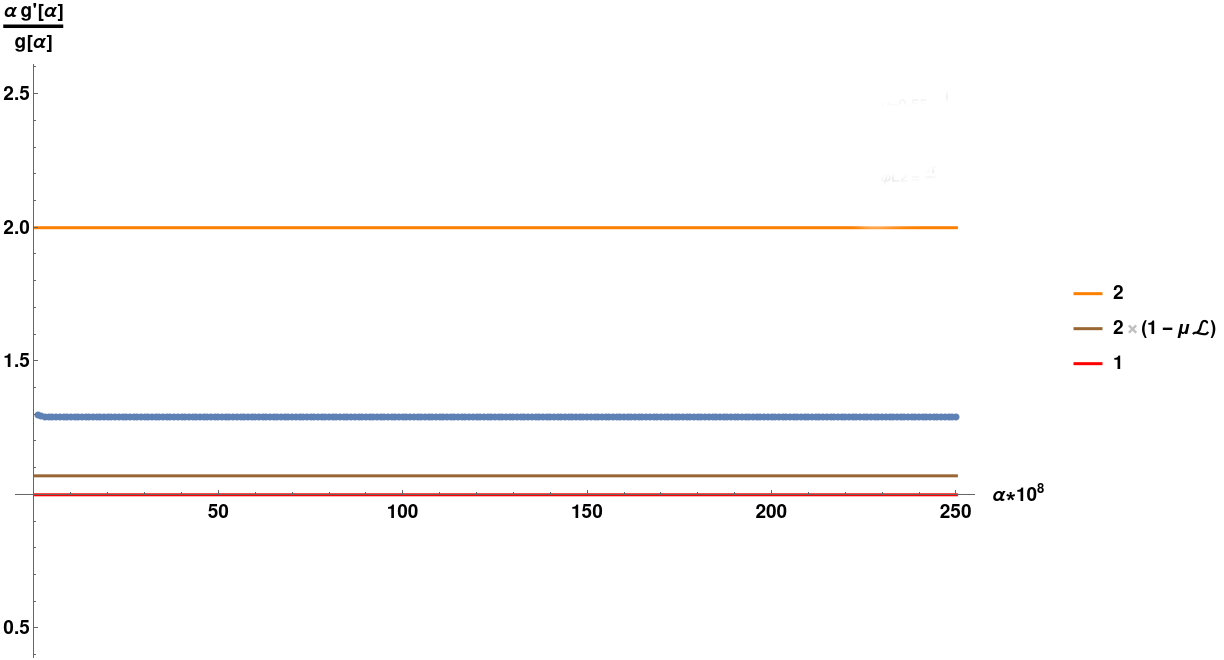}
\includegraphics[scale=0.25]{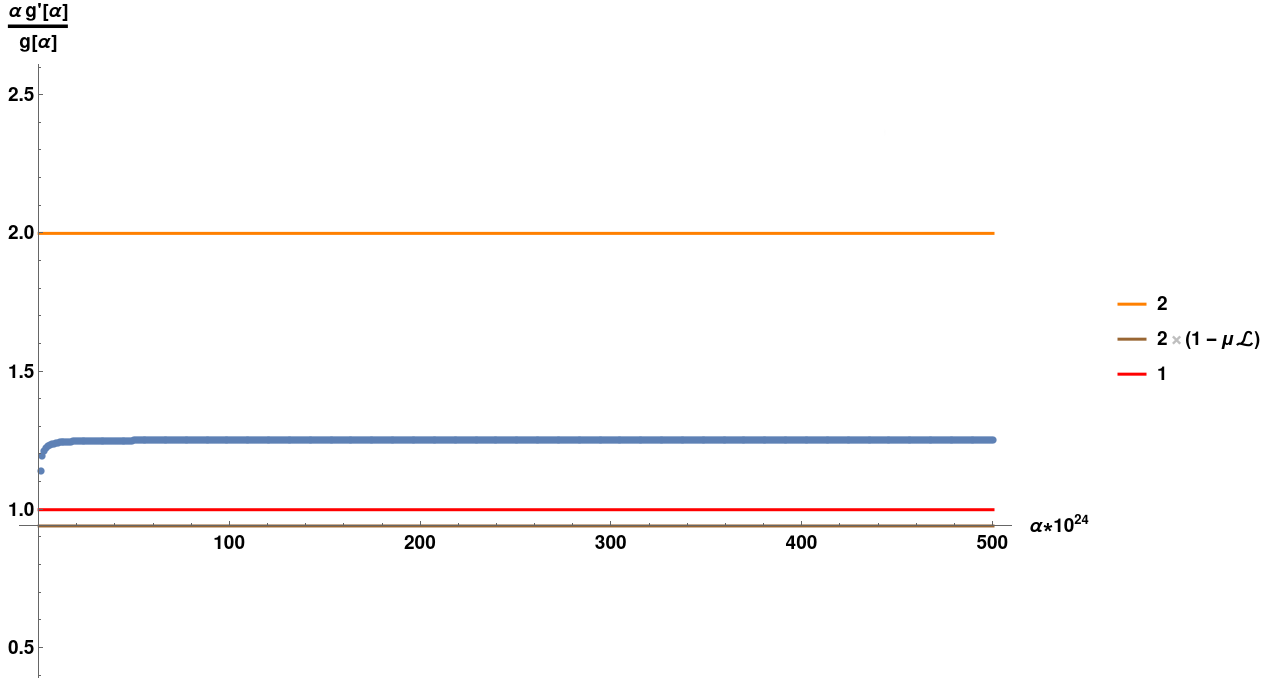}
\caption{Plots in blue for different values of $\mu$ and $\mathcal{L}$ $s.t.$ $1>\mathcal{L}\geq 2\mu/(1+\mu^2)$. Here the $y$-axis measures $2\lambda_L/\kappa$, therefore the line at $y=2$ (orange) implies $\lambda_L=\kappa$ while $y=1$ implies $\lambda_L=\frac{\kappa}{2}$ (red) and at $y=2(1-\mu\mathcal{L})$ implies $\lambda_L=2\pi/\beta$ (brown). The ranges of $\alpha$ on the $x$-axis is $s.t.$ $S_{AB}\leq S_A+S_B$ and hence are different for different values of parameters. From top left clockwise: $\rpl=35,\mu=0.45,\mathcal{L}=0.75,\phi_{L_1}=\pi,\phi_{L_2}=\pi/4$; $\,\,\,\,\rpl=25,\mu=0.5,\mathcal{L}=0.8,\phi_{L_1}=\pi,\phi_{L_2}=\pi/2$;  $\,\,\,\,\rpl=30,\mu=0.6,\mathcal{L}=0.88,\phi_{L_1}=\pi,\phi_{L_2}=\pi/2$; $\,\,\,\,\rpl=35,\mu=0.55,\mathcal{L}=0.84,z_{L_1}=\pi,\phi_{L_2}=\pi/4$.}
\label{F1}
\end{figure}
\\\\
To analyse cases with $\mathcal{L}<1$ we would have to plot $S_{AB}$ numerically as analytic solutions to finding minima of \eqref{func_minima} are not possible. For this we note that $S_{AB}=\log(1 + g(\alpha))$ where $g(\alpha)$ is some unknown function to be analysed numerically $s.t.$ $g(0)=0$ as the geodesics traversing the black holes would have zero lengths without the perturbation. $g(\alpha)$ in particular is given by
\bea
g(\alpha)&=&k(z_{L1},\alpha)k(z_{L_2},\alpha) -1\cr&&\cr
k(z_L,z)&=&{\rm Min}_z[\cosh(\rpl(z_{L}-z))+e^{\rpl\tfrac{\mu-\mathcal{L}}{1-\mu\mathcal{L}}(z_{L}-z)}\, \frac{\alpha e^{\kappa \mathcal{L}z_{L}}}{2}]
\eea
Since mutual information is given by 
\be
I(A:B)=\frac{l}{2G_N}\left[S_A+S_B-\log(1+g(\alpha))\right]
\ee
therefore $\lambda_L$ would be given by the highest power of $\alpha$ in the function $g(\alpha)$. For example: for $g(\alpha)\sim \alpha^2$ implies $\lambda_L=\kappa$ as in the case for $\mu=0=\mathcal{L}$, while $g(\alpha)\sim\alpha$ implies $\lambda=\kappa/2$ as in $\mathcal{L}=1$ case.  Plotting $\alpha\,\partial_\alpha \log g$ we find for certain cases (Figure \ref{F1}) we find 
\bea
&&\frac{\kappa}{2}<\frac{2\pi}{\beta}<\lambda_L<\kappa\cr&&\cr
{\rm or}&&\frac{2\pi}{\beta}<\frac{\kappa}{2}<\lambda_L<\kappa
\eea 
$i.e.$ we do find that the Lyapunov index is greater than the temperature but bounded by $\kappa$.
There are also values of parameters which imply $\lambda_L<2\pi/\beta$, in Figure \ref{F1} we have only shown the plots for cases where it is greater than the temperature. 
\\\\  
To determine the scrambling time one also needs to know how the coefficient of the exponentially growing term depends on the black hole parameters. More specifically it would be required to know it's dependence on the black hole's entropy.
Since this coefficient $i.e.$ that of the highest power of $\alpha$ in $g(\alpha)$ can't be determined as a function of black hole parameters we cannot estimate how the scrambling time for $I(A:B)$ scales with $t_0$ for generic $\mathcal{L}$.
\section{Conclusion and Discussion}
We have analysed the effect of rotating shock-waves on mutual information  between the left and right CFTs in a TFD state dual to a generic eternal BTZ black hole. We learn that in general the Lyapunov index is bounded by the blueshift seen along the shock-wave which is the greater of the 2 temperatures of the dual $CFT_2$. For the case with $\mathcal{L}=1$ we are able to solve for the mutual information at late times analytically and we find that $\lambda_L=\kappa/2$, which for $\mu>1/2$ is greater than the temperature of the black hole. For this case we also see that the scrambling time is indeed given by $t_*\sim\frac{2}{\kappa}\log S$. This conclusively shows that at large time scales the disruption of mutual information can be governed by the temperatures greater than that of the system's temperature depending upon the perturbation. We have    also plotted the values of Lyapunov index where analytical solutions are not possible and find that $\lambda_L$ is bounded by $\kappa$ and not $\kappa/2$ or the black hole's temperature.
\\\\
Our results seem to obey the bound \eqref{Halder_bound} computed in \cite{Halder:2019ric} which is applicable to rotating geometries in holographic theories as the MSS bound \cite{Maldacena:2015waa} holds only for non rotating geometries. Our results are also consistent with the similar CFT$_2$ computations done in \cite{Stikonas:2018ane} as explained in subsection-2.2.  Note that although the results of \cite{Reynolds:2016pmi} can be regarded as a special case of our analysis with $\mathcal{L}=0$, the technique used there to obtain the shockwave solution seem to be at odds with our analysis $c.f.$ subsection 2.2. The 4pt. OTOCs computed for rotating geometries in \cite{Mezei:2019dfv,Craps:2020ahu,Craps:2021bmz} find that the scrambling time is governed by smaller of the 2 temperatures, this does not seem to contradict our findings here as entanglement entropy is inherently non-local in its nature.   
\\\\
It is nonetheless interesting to juxtapose the results  of this paper  with earlier works which have computed 4pt. OTOCs in extremal \cite{Mezei:2019dfv} and non-extremal BTZ \cite{Craps:2021bmz,Craps:2020ahu} for longer time scales. The authors in these works found that although momentarily $\lambda_L$ is greater of the 2 temperatures of the dual CFT$_2$, the OTOC decreases at a rate primarily controlled by the smaller of the 2 temperatures. It is important to point out that the analytic arguments used in \cite{Maldacena:2015waa} for large $N$ thermal QFT and it's generalization to include chemical potential in \cite{Halder:2019ric} do not in any way restrict the behaviour of $\lambda_L$ averaged over large time scales but only the instantaneous value of $\lambda_L$. In other words there is no universal bound on the average of $\lambda_L$ other than the one imposed by it instantaneous value. The results summarized above indicate that the long time behaviour of chaotic dynamics crucially depends upon the type of perturbation used to instigate the change along with the  quantity being measured. Pole-skipping is yet another measure of chaotic phenomena \cite{Natsuume:2019vcv,Natsuume:2019sfp,Natsuume:2020snz,Blake:2021hjj,Blake:2021wqj,Blake:2018leo,Liu:2020yaf,Choi:2020tdj} and it would be interesting see if it can be used to deduce information about chaos in higher dimensional black holes, $c.f.$ \cite{Blake:2021hjj} for attempts at understanding pole-skipping and OTOCs in Kerr-$AdS_4$. We also point out the pole-skipping computed for rotating BTZ in \cite{Liu:2020yaf} does see a $\lambda_{L\pm}=2\pi/\beta_{\pm}$ consistent with \cite{Poojary:2018eszz,Jahnke:2019gxr}. 
\\\\
The JT model describes an effective gravitational theory for near horizon dynamics of large near extremal black holes \cite{Jensen:2016pah,Maldacena:2016upp} which captures the chaotic behaviour characterized by $\lambda_L=2\pi/\beta$. This being  obtained from dimensional reduction to two dimensions from higher dimensional near extremal black holes seems  to be sensitive to only those chaotic modes which correspond to $\lambda_L=2\pi/\beta$. For the case of BTZ this corresponds to the lower of the 2 temperatures of the CFT$_2$. The JT model as described in \cite{Jensen:2016pah,Maldacena:2016upp}  cannot explain the chaotic behaviour characterized by the larger of the 2 CFT$_2$ temperatures especially since this temperature does not vanish at extremality. For black holes in $AdS_{d>3}$ there has not been any direct calculation of scrambling or the Lyapunov index except for $via$ the JT model \cite{Nayak:2018qej,Moitra:2018jqs,Moitra:2019bub}.  As evident from our analysis in the present article, if the intuition that $\lambda_L$ is generally bounded by the blueshift seen by an in-falling null trajectory  holds true even in higher dimensions then we can expect that $\lambda_L\leq 2\pi/\beta$ for RN and Schwarzchild black holes. It would be interesting to understand a similar computation for scrambling of in fallen information in Kerr-$AdS$ which is a work we leave for the near future. The JT model arising in the near horizon region of near extremal Kerr in 4 and 5 dimensions were investigated in \cite{Castro:2018ffi} and later in \cite{Castro:2021csm,Castro:2021fhc} and were found to have many interesting non-universal features. The Lyapunov index was also computed in 3 dimensional flat geometries with a cosmological horizon both by using shock waves and the dual Galilean CFT and was found to be the temperature associated with the horizon \cite{Bagchi:2021qfe}.  It would be also interesting to investigate a possible effective lower dimensional gravitational theory like the JT model explaining these chaotic modes which at least for the case of rotating BTZ gives rise to fast scrambling even at zero temperature. The butterfly velocity is the spread of this disruption in the transverse coordinates and has interesting physics contained in it $c.f.$ \cite{Mezei:2019dfv,Alishahiha:2016cjk}. It would be interesting to probe these  issues for rotating geometries in a similar manner. 
\\\\
Recently sub-leading bounds on chaos were investigated in \cite{Kundu:2021qcx,Kundu:2021mex} where if the Lyapunov exponent- as defined by first sub-leading correction in $G_N$ (or $1/c$), saturates the chaos bound then analyticity of OTOCs places bounds on  similar exponents occurring in the subsequent sub-subleading orders.  It would be interesting to understand these bounds in terms of mutual information wherein the subleading corrections to the RT and the HRT surfaces are given by the generalized  quantum extremal surfaces \cite{Engelhardt:2014gca,Faulkner:2013ana}. These surfaces have lead to the resolution of the entropic black hole information paradox \cite{Penington:2019npb,Almheiri:2019psf,Lewkowycz:2013nqa}. It would be very interesting to understand how a shockwave triggered by infinitesimal in-falling quanta at very late times effects the generalized quantum extremal surface.  

\section*{Acknowledgements} The authors would like to thank Sukruti Bansal for collaboration in an early stage. The authors would also like to thank Arnab Kundu for his comments on a draft of this work. RP would like to thank Daniel Grumiller and Arnab Kundu for discussions on various aspects of this project. RP is supported by the Lise Meitner project FWF M-2883 N. 
\appendix
\section{Single particle back-reaction}
Here we note the back reaction due to a single particle for completeness. 
The back reaction in the transverse direction for a single null particle  is captured by $f(z)$ obeying \eqref{back_reaction_f}. Here the differential operator $\mathcal{D}$ is of the form 
\be
A\partial_z^2 +B\partial_z +C
\ee
where the coefficients are given in terms of finite quantities at the horizon $\{\tfrac{F}{UV},\left(\tfrac{F}{UV}\right)',h,\tfrac{h_1}{UV},\kappa\}$.
\bea
&&A=-\left(\tfrac{F}{\kappa^2UV}\right)\left(\tfrac{1}{2h}\right),\,\,B=\tfrac{h_1}{\kappa UV},\,\,C=\tfrac{2\kappa^2 UV}{F}\left[\left(\tfrac{F}{\kappa^2UV}\right)^2-4h h_1^2+\left(\tfrac{F}{\kappa^2UV}\right)'\right]\cr&&\cr
{\rm where\, \,at} \,\,\rpl: && \left(\tfrac{F}{\kappa^2UV}\right)=-4,\,\,h=rp^2,\,\,\tfrac{h_1}{\kappa UV}=\tfrac{-4(\mu-\mathcal{L})}{\rpl(1-\mu\mathcal{L})^3},\cr&&\cr
&&\left(\tfrac{F}{\kappa^2UV}\right)'=\tfrac{16\kappa^2F''(\rpl)}{F'(\rpl)^2}=8(1-\mu\mathcal{L})(1-4\mathcal{L}^2)
\eea
As $z=(\phi-\mu t)$, at fixed $t$, $z$ has a periodicity of $2\pi$. Therefore we express the solution to \eqref{back_reaction_f}  as
\bea
f(x)&=&\sum_{n=-\infty}^\infty\frac{e^{in(z-z_p)}}{-An^2+iBn+C}=(2\pi i)\oint \frac{e^{w(z-zp)}\,\,2\pi\,\,dw}{(Aw^2+Bw+C)(e^{2\pi w}-1)}\cr&&\cr
&=& \frac{2\pi}{(w_+-w_-)}\left(-\frac{e^{w_+(z-z_p)}}{(e^{2\pi w_+}-1)}+\frac{e^{w_-(z-z_p)}}{(e^{2\pi w_-}-1)}\right)\cr&&\cr
{\rm with}\,\,\,\,  w_\pm&=&\rpl\left(\tfrac{\mu-\mathcal{L}}{\lambda^3}\right)\pm\rpl\sqrt{2+8\mathcal{L}^3\mu+\mathcal{L}^2\left(\tfrac{5}{\lambda^6}-8\right)+2\mathcal{L}\mu\left(1-\tfrac{5}{\lambda^6}\right)+5\tfrac{\mu^2}{\lambda^6}}\cr&&\cr
&=&\pm\sqrt{2}\rpl, \,\,\,\,\,{\rm for}\,\,\,{\mu=\mathcal{L}=0}
\label{f_sol}
\eea
where $\lambda=({1-\mu\mathcal{L}})$, and $w_\pm$ solve $Aw^2+Bw+C=0$. It is important to note that this back-reaction in the transverse direction at late times is again independent of $U_0$.
\section{RN-AdS$_4$}
\label{RN}
One can similarly analyse affine coordinates for rotating shockwaves for charged static black holes in higher dimensions. We note here the resulting $\kappa$ close to the horizon of a Reissner-Nordstr\"{o}m black hole in $AdS_4$ for a generic null shock-wave with arbitrary $\mathcal{L}$ as in the previous case. The metric for RN-$AdS_4$ in Boyer Lindquist coordinates takes the form
\bea
ds^2_{RN}&=&\frac{dr^2}{f(r)^2}-dt^2f(r)^2+r^2(d\theta^2+\sin^2\theta \,d\phi^2)^2\cr&&\cr
{\rm where}\,\,\,
f(r)&=& 1-\frac{2M}{r}+\frac{4\pi Q^2}{r^2}+r^2.
\label{RN_AdS_4}
\eea  
We then find in-falling and out going null geodesic vector fields $\xi_\pm\cdot\partial$ by solving \eqref{null_geodesic_0} for the above metric. Here we take $\xi^\theta$ component to be zero as the geometry is spherically symmetric. We then write the metric in terms of the line elements along these null vector field pairs like \eqref{LC_BTZ} as
\bea
ds^2_{RN}=F(r)dudv+h(r)\sin^2\theta(d\phi+h_1(r)dt)^2+r^2d\theta^2
\eea
Defining $\mathcal{K}$ as in \eqref{kappa} and evaluating it at the outer horizon  we find
\be
\kappa=\left.\mathcal{K}\right\vert_{\rpl}=\left|\tfrac{1}{2}\xi_\pm.\partial F\right\vert_{\rpl}=\frac{2\pi}{\beta}
\ee
Therefore we find that rotating null geodesics see the same extrinsic curvature at the horizon as  non-rotating ones $i.e.$ the temperature of the RN-$AdS_4$. One can easily see that the analysis is essentially dimension independent.
\section{Kerr-AdS$_4$}
\label{Kerr}
The above analysis can be similarly repeated for Kerr-$AdS_4$ by taking into account an additional conserved charge $Q$ - the Carter's constant, associated with the Killing-Yano tensor for a null geodesic
\be
\xi^\mu K_{\mu\nu}\xi^\nu=Q.
\label{Killing_Yano_charge}
\ee
We mention here that the $\kappa$ computed at the outer horizon is independent of $Q$ and the polar coordinate $\theta$ but like in the rotating BTZ case depends on $\mathcal{L}$ and can be greater than its temperature $2\pi/\beta$. We postpone the analysis of perturbation of mutual information in Kerr-$AdS_4$ for the near future \cite{Malvimat:2022ongoing}

\bibliographystyle{JHEP.bst}
\bibliography{bulk_syk_soft_modes.bib}
\end{document}